# Lock-in-detection dual-comb spectroscopy

Hidenori Koresawa,[1,2] Kyuki Shibuya,[2,3] Takeo Minamikawa,[2-4] Akifumi Asahara,[2,5] Ryo Oe,[1,2] Takahiko Mizuno,[2-4] Masatomo Yamagiwa,[2,3] Yasuhiro Mizutani,[2,6] Tetsuo Iwata,[2,3] Hirotsugu Yamamoto,[2,7] Kaoru Minoshima,[2,5] and Takeshi Yasui[2-4,*]

[1]*Graduate School of Advanced Technology and Science, Tokushima University, 2-1, Minami-Josanjima, Tokushima, Tokushima 770-8506, Japan*

[2]*JST, ERATO, MINOSHIMA Intelligent Optical Synthesizer Project, 2-1, Minami-Josanjima, Tokushima, Tokushima 770-8506, Japan*

[3]*Graduate School of Technology, Industrial and Social Sciences, Tokushima University, 2-1, Minami-Josanjima, Tokushima, Tokushima 770-8506, Japan*

[4]*Institute of Post-LED Photonics, Tokushima University, 2-1, Minami-Josanjima, Tokushima, Tokushima 770-8506, Japan*

[5]*Graduate School of Informatics and Engineering, The University of Electro-Communications, 1-5-1 Chofugaoka, Chofu, Tokyo 182-8585, Japan*

[6]*Graduate School of Engineering, Osaka University, 2-1, Yamadaoka, Suita, Osaka 565-0871, Japan*

*[7]Center for Optical Research and Education, Utsunomiya University, 7-1-2, Yoto, Utsunomiya, Tochigi 321-8585, Japan*

*\*yasui.takeshi@tokushima-u.ac.jp*

**Abstract**

Dual-comb spectroscopy (DCS) is useful for gas spectroscopy due to high potential of optical frequency comb (OFC). However, fast Fourier transform (FFT) calculation of a huge amount of temporal data spends significantly longer time than the acquisition time of an interferogram. In this article, we demonstrate frequency-domain DCS by a combination of DCS with lock-in detection, namely LID-DCS. LID-DCS directly extracts an arbitrary OFC mode from a vast number of OFC modes without the need for FFT calculation. Usefulness of LID-DCS is demonstrated in rapid monitoring of transient signal change and spectroscopy of hydrogen cyanide gas.

# 1. Introduction

Recent advances in optical frequency comb (OFC) [1-3] enable us to benefit from a group of a vast number of phase-locked narrow-linewidth continuous-wave (CW) lights with a constant frequency spacing $f_{rep}$ (typically, 50 to 100 MHz) over a broad spectral range. The inherent mode-locking nature and active laser control make it possible to use the OFC as an optical frequency ruler traceable to a microwave or radio-frequency (RF) frequency standard. To fully utilize both its narrow spectral linewidth and broadband spectral coverage for broadband spectroscopy, it is essential to acquire the mode-resolved OFC spectrum. Fourier transform spectroscopy [4] can be used for this purpose by using a long mechanical scanning of a reference arm (typically, sub-meter to a few meters length) [5]; however, such mechanical scanning hampers rapid data acquisition. Virtually-imaged-phased-array (VIPA) spectroscopy [6, 7, 8] is a promising method for rapid acquisition of mode-resolved OFC spectrum. By spatially developing OFC spectrum with a combination of VIPA and a diffraction grating [9, 10], the mode-resolved OFC spectrum can be acquired all at once as a two-dimensional (2D) spectrograph by a camera without the need for mechanical scanning. However, VIPA spectroscopy is limited for OFCs with $f_{rep}$ larger than a few GHz due to its spectral resolving power.

Recently, dual-comb spectroscopy (DCS) [11-14] has appeared as a technique for acquiring the mode-resolved OFC spectrum via its replica in RF regions, namely RF combs, by using dual OFCs with slightly mismatched frequency spacing (= $f_{rep1}$ and $f_{rep2}$). Due to its rapid, precise, and accurate acquisition of the spectrum, DCS has found many applications in optical frequency metrology; examples include gas spectroscopy [15], gas thermometry [16], solid spectroscopy [17], spectroscopic ellipsometry [18], hyper-spectral imaging [19], and coherent Raman imaging [20]. Among them, gas spectroscopy is one interesting application because DCS-based gas analysis has several advantages over conventional gas analysis including gas chromatography: real-time data acquisition, simultaneous analysis of multiple gasses, and no need for sample preparation. For example, the broad-band DCS covering from 158 to 300 THz, corresponding to 1.0 to 1.9 μm, has been effectively applied for simultaneous analysis of acetylene, methane, and water vapor [21]. Also, DCS has been used for monitoring of atmospheric gas [22, 23] and gas turbine exhaust [24]. Furthermore, such DCS has been extended to the mid-infrared region [25] and even the terahertz (THz) region [26, 27].

In usual DCS, after a temporal waveform of a single interferogram or consecutive interferograms was acquired in time domain, mode-resolved OFC spectrum is obtained by fast Fourier transform (FFT) calculation of the acquired temporal waveform. However, FFT calculation consumes time due to a huge amount of temporal data for the mode-resolved OFC spectra. Due to this FFT calculation, the actual measurement rate significantly decreases even though the acquisition rate of temporal waveform can be increased up to a difference of $f_{rep}$ between dual OFCs (= $\Delta f_{rep} = f_{rep2} - f_{rep1}$); it will hamper monitoring of transient signal change. Low duty factor in the ultra-discrete tooth-like spectrum of OFC is another practical limitation of DCS. For example, when the mode-resolved OFC spectrum is measured by a spectral resolution of $f_{rep}/100$, the mode linewidth of the measured OFC spectrum is decreased down to $f_{rep}/100$. However, spectral data points except mode peaks fall in gap regions between OFC modes; only 1 % of the spectral data points gives the information on the signal of mode peaks, and the remaining 99 % of them gives no information due to noise region without OFC modes. Furthermore, in the case of gas spectroscopy, the absorption lines of the gas molecule are localized at specific spectral region [21]; it is not always necessary to acquire the whole spectral range of OFC, and only the spectral information at the absorption lines is sufficient for simple analysis. Therefore, the spectral analysis with minimum required spectral information is greatly desired for efficient and fast DCS.

One possible method of overcoming these limitations is the frequency-domain acquisition in DCS by use of lock-in detection (LID). Since the RF comb has a highly stable, discrete spectrum in frequency domain, one can extract only a specific RF comb mode by selection of a LID reference frequency; simultaneously, other unnecessary RF comb modes and gap data points can be rejected. This leads to the great reduction of the data size. Also, since the LID is based on the frequency-domain measurement, it needs no FFT calculation to obtain the spectral information, enabling fast processing in DCS. Its acquisition time is dependent on a LID time constant independently of $\Delta f_{rep}$. Combination of LID with DCS, namely LID-DCS, has been successfully demonstrated in DCS-based distance measurement, in which the optical phase of the specific RF comb mode was measured by LID [28, 29]. However, there are no attempts to apply LID-DCS for gas spectroscopy requiring the optical amplitude measurement of specific RF comb mode.

In this paper, we evaluate the basic performance of LID-DCS by comparing with usual DCS from viewpoint of net measurement time and signal-to-nose ratio (SNR). We further demonstrate use of LID-DCS for spectroscopy of hydrogen cyanide gas.

## 2. Experimental setup

Figure 1(a) shows a principle of operation in LID-DCS. In the frequency-domain description of DCS, two OFCs with a slightly different repetition frequency (signal OFC, mode spacing = $f_{rep1}$; local OFC, mode spacing = $f_{rep2} = f_{rep1} + \Delta f_{rep}$) generates a secondary frequency comb in RF region, namely RF comb (mode spacing = $\Delta f_{rep}$), via the multi-frequency heterodyning interference between them. In usual DCS, the RF comb are acquired as an RF interferogram in time domain and then are obtained as the mode-resolved spectrum by FFT calculation of the RF interferogram. In LID-DCS, a lock-in amplifier (LIA) enables us to acquire both amplitude and phase of a frequency signal synchronized with a LID reference-frequency signal. Therefore, one can select an arbitrary mode from the mode-resolved RF comb spectrum without the need for FFT by tuning the LID reference frequency to coincide with a target RF-comb-mode frequency.

Figure 1(b) shows an experimental setup of LID-DCS. Two mode-locked erbium-doped fiber combs (OCLS-HSC-D100-TKSM, Neoark Co., Japan; center wavelength = 1560 nm, spectral bandwidth = 50 nm; signal OFC, carrier-envelope-offset frequency = $f_{ceo1}$ = 10.5 MHz, $f_{rep1}$ = 100.000188 MHz ; local OFC, $f_{ceo2}$ = 10.5 MHz, $f_{rep2}$ = 99.999976 MHz; $\Delta f_{rep}$ = $f_{rep2} - f_{rep1}$ = 212 Hz) were used for light sources in LID-DCS. We used a rubidium frequency standard (Rb-FS, Stanford Research Systems, Inc., FS725; frequency = 10 MHz, accuracy = 5×10$^{-11}$; instability = 2×10$^{-11}$ at 1 s) for a frequency reference in these dual OFCs. The local OFC, equipped with an intra-cavity electro-optical modulator for laser control, was tightly and coherently locked to the signal OFC with a frequency offset using a narrow-linewidth continuous-wave (CW) laser (CWL, Redfern Integrated Optics, Inc., Santa Clara, California, USA, PLANEX; center wavelength = 1550 nm; FWHM < 2.0 kHz) for an intermediate laser [16-18]. Polarization of the signal OFC light and the local OFC light was aligned at the vertical direction by use pairs of a quarter waveplate (λ/4) and a half waveplate (λ/2). After spatially overlapping of them for optical interference by a beam splitter (BS), the dual OFC lights passed through a band-pass filter (BPF, pass band = 1550 ± 10 nm) for bandwidth reduction and another λ/2 for polarization rotation by 45º. Then, the dual OFC lights were split for a signal light and a reference light by a polarization beam splitter (PBS). A sample was placed into the optical path of the signal light. The RF combs of the signal light and the reference light, namely signal RF comb and reference one, were respectively detected by a pair of photodetectors (PDs, Thorlabs, PDA10CF-EC; wavelength = 800–1700 nm; bandwidth <

150MHz). We extract an arbitrary comb mode from the signal RF comb by a radio-frequency LIA (RF-LIA1, Stanford Research Systems, SR844; frequency range = 25kHz ~ 200MHz, time constant = no or 100 μs to 30 ks). We further extracted the same-order comb mode of the reference RF comb for a reference to compensate the common-mode fluctuation in amplitude, arising from dual OFCs such by the fluctuation of temperature, air flow and so on, by use of another RF-LIA (RF-LIA2, Stanford Research Systems, SR844) in real-time. Then, we calculated amplitude ratio between them as a normalized amplitude spectrum. LID reference-frequency signals for RF-LIA1 and RF-LIA2 were generated from a RF waveform generator (RF-WG, Keysight Technologies, 33510B, frequency range < 20MHz). Since dual OFCs and the RF-WG share the same Rb-FS for the common frequency reference, the LID reference-frequency signal can be synchronized with the arbitrary RF comb mode.

For comparison with LID-DCS, we performed usual DCS using the same optical setup except the reference light in Fig. 1(b). Detail of its experimental setup is given elsewhere [18, 19]. The detected electrical signal was acquired using a digitizer (National Instruments Corp., NI PXIe-5122; resolution = 14 bit). The sampling clock signal was synchronized with $f_{rep2}$. We made an FFT calculation program to obtain full spectrum of amplitude and phase in OFC with LabView2017 (National Instruments Corp., 64 bit) and performed it in a computer (National Instruments Corp., PXIe-8840, Intel Core i7, Processor base frequency = 2.60 GHz, Cache = 6 MB smart cache, RAM = 8GB).

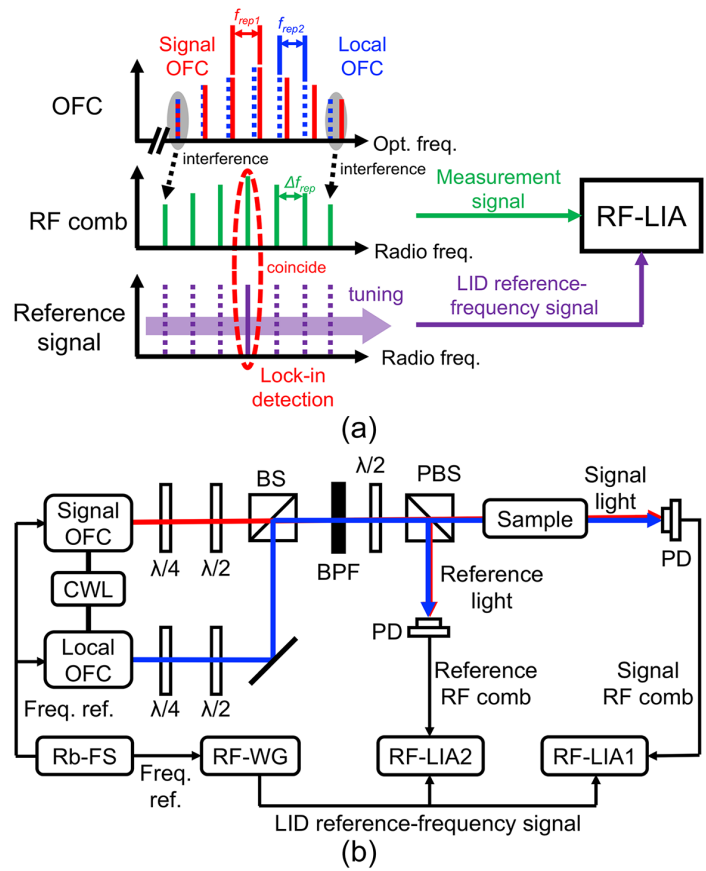

Fig. 1. (a) Principle of operation and (b) experimental setup of LID-DCS. Signal and local OFCs, signal and local optical frequency combs; CWL, narrow-linewidth CW laser; Rb-FS, rubidium frequency standard; λ/4, a quarter waveplate; λ/2, half waveplate; BS, beam splitter; BPF, 1550 ± 10 nm band-pass filter; PBS, polarization beam splitter; PD, photodetector; RF-LIA1 and RF-LIA2, radio-frequency lock-in amplifiers; RF-WG, RF waveform generator.

# 3. Results

*3.1 Performance evaluation of LID-DCS and DCS*

We first investigated a relation between the number of measured spectra and the net measurement time under the same optical frequency resolution. The net measurement time was defined as an acquisition time of accumulated signal for LID-DCS and a sum of acquisition time of a single interferogram or consecutive interferograms and FFT calculation time for DCS. Red and blue lines in Fig. 2(a) respectively show a relation between the number of measured spectra and the net measurement time for LID-DCS (optical spectral resolution = 10.8 MHz, LID time constant = 3 ms, number of signal accumulation = 14, net measurement time = 42 ms/a single spectral point) and DCS (optical spectral resolution = 11.1 MHz, number of consecutive interferograms = 9, time window size = $9/f_{rep1}$ = 90 ns, sampling time interval = $1/f_{rep2}-1/f_{rep1}$ = 21.2 fs, number of temporal data = 4,245,282, acquisition time of interferogram = $9/\Delta f_{rep}$ = 42 ms/data, number of interferogram accumulation = 1, FFT calculation time = 3.08 s/data, net measurement time = 3.12 s/a single spectral point). The net measurement time of LID-DCS was made to coincide with the acquisition time of consecutive interferograms in DCS by accumulating 14 data in LID-DCS. In LID-DCS, the net measurement time is determined by the LID time constant and the number of signal accumulation; in DCS, most of the net measurement time is occupied by the FFT calculation time rather than the acquisition time of interferogram. This comparison indicated an advantage of LID-DCS over DCS in the net measurement time.

We next measured a fluctuation of spectral amplitude in LID-DCS and DCS on a single spectral point when the net measurement time in LID-DCS (= 42 ms/a single spectral point) was set to be equal to the acquisition time of 9 consecutive interferogram in DCS (= 42 ms/data). Figures 2(b) and 2(c) compare a fluctuation of spectral amplitude at 193.554964 THz between LID-DCS (optical spectral resolution = 10.8 MHz, LID time constant = 3 ms, number of signal accumulation = 14, net measurement time = 42 ms/a single spectral point) and DCS (optical spectral resolution = 11.1 MHz, number of consecutive interferograms = 9, time window size = 90 ns, sampling time interval = 21.2 fs, number of temporal data = 4,716,980, acquisition time of interferogram = 42 ms/data, number of interferogram accumulation = 1, FFT calculation time = 3.08 s/data, net measurement time = 3.12 s/a single spectral point) with respect to the number of measured spectra. Similar fluctuation of spectral amplitude was observed in both. When we defined SNR as a ratio of the mean to the standard deviation in spectral amplitude, SNR in them is significantly similar to each other: 30.6 for LID-DCS and 31.9 for DCS. Therefore, use of LID in DCS does not contribute to negative effect in SNR. Although the total number of measured spectra was equal to each other in Figs. 2(b) and 2(c), there is a large difference of the net measurement time between them if the total number of measured spectra is

-12-

converted into the total net measurement time based on Fig. 2(a): 2.8 s for LID-DCS and 206 s for DCS.

Considering the results in Figs. 2(a), 2(b), and 2(c), the LID-DCS can reduce the net measurement time while maintaining the SNR and the spectral resolution similar to DCS. Figure 2(d) compares SNR of spectral amplitude of a single spectral point at 193.559964 THz with respect to the net measurement time between LID-DCS (see red circle) and DCS (see blue circle). We here adjusted the optical resolution of DCS while maintaining the constant optical resolution of LID-DCS to match the net measurement time of LID-DCS and that of DCS (see red and blue triangles). The linear relation was confirmed between SNR and net measurement time in both methods. Starting point of the slope in LID-DCS (see red line) was significantly higher SNR and shorter net measurement time than that in DCS because of no FFT calculation. The slope coefficient was determined to be 0.37 for LID-DCS and 0.57 for DCS, respectively. In LID-DCS, while the residual timing jitter between dual OFCs somewhat fluctuates frequency of RF comb modes, the LID reference frequency is always fixed at a constant value. Lock-in detection of such frequency-fluctuated signal at a fixed frequency makes the LID-DCS sensitive to the timing jitter and hence limits the slope coefficient. In the case of DCS, we applied the phase compensation for the RF interferogram, in which the phase of the RF interferogram was preset to null for every interferogram by the self-triggering of the RF interferogram in the acquisition of temporal waveform, making DCS robust to the residual timing jitter. Difference of slope coefficient between LID-DCS and DCS is mainly due to these effects.

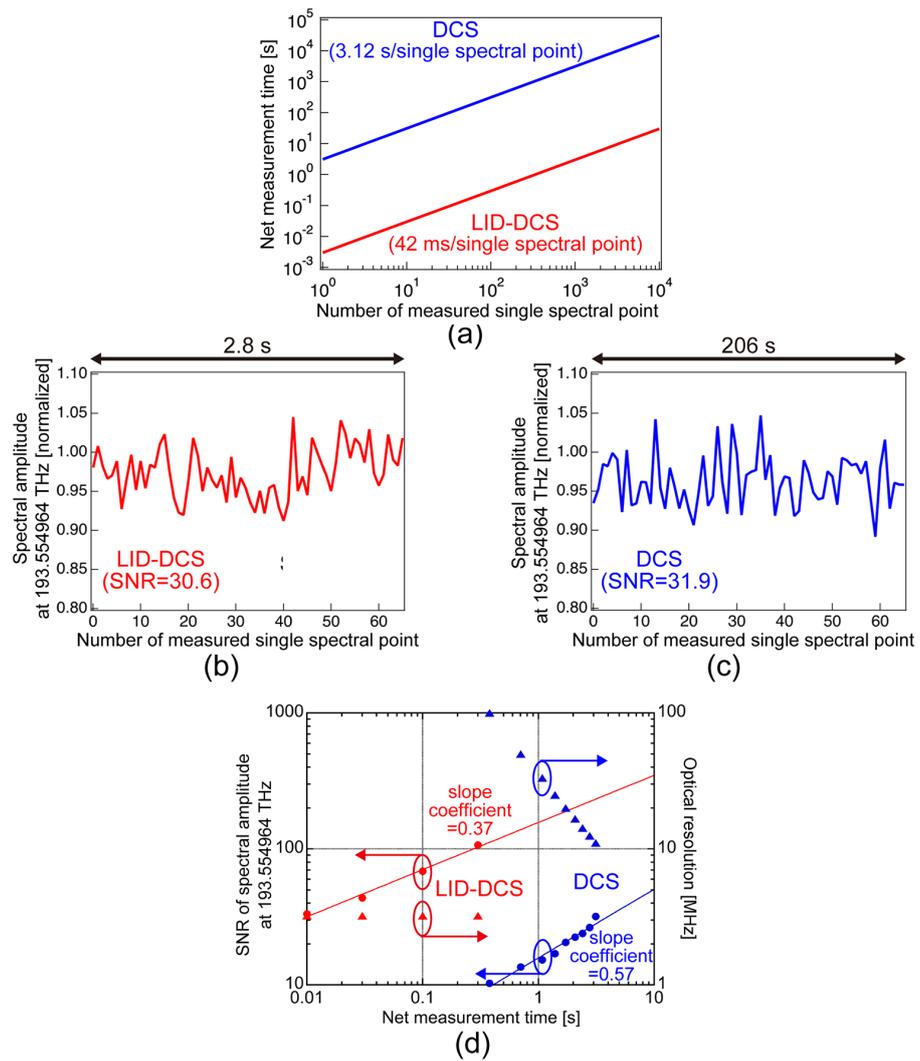

Fig. 2. Performance evaluation of LID-DCS and DCS. (a) Relation between the number of measured spectra and the net measurement time. Fluctuation of spectral

amplitude in (b) LID-DCS and (c) DCS. (d) Comparison of SNR of spectral amplitude and optical resolution between LID-DCS and DCS.

*3.2 Temporal response of LID-DCS and DCS*

We next evaluated the temporal response of LID-DCS and DCS when the intensity of the measured signal light was transiently fluctuated. To this end, we chopped the optical beam with a glass plate (BK7, thickness = 1mm), leading to a transient change in the optical intensity. Figure 3 shows the temporal response of the spectral amplitude at 193.554964 THz for (a) LID-DCS (optical spectral resolution = 1.1 MHz, LID time constant = 30 ms, number of signal accumulation = 1, net acquisition time = 30 ms/a single spectral point) and (b) DCS (optical spectral resolution = 100 MHz, number of consecutive interferograms = 1, time window size = $1/f_{rep1}$ = 10 ns, sampling time interval = 21.2 fs, number of temporal data = 471,698, number of signal accumulation = 6, acquisition time of interferogram = $1/\Delta f_{rep}$ = 4.7 ms/data, FFT calculation time = 0.373 s/data, net acquisition time = 0.401 s/a single spectral point) with respect to the elapsed time. In this experiment, optical resolution of DCS was 100-times worse than that of LID-DCS to maintain the real-time capability of FFT calculation in DCS by reducing the data size of temporal waveform of the RF interferogram. Nevertheless, the DCS has less discrete sampling points and could not respond to such transient fluctuation sensitively; in contrast, the LID-DCS well responds to the fluctuation by sufficient number of sampling points. Therefore, the LID-DCS will be more powerful than DCS for monitoring of transient signal change, such as gas concentration measurement under air turbulence.

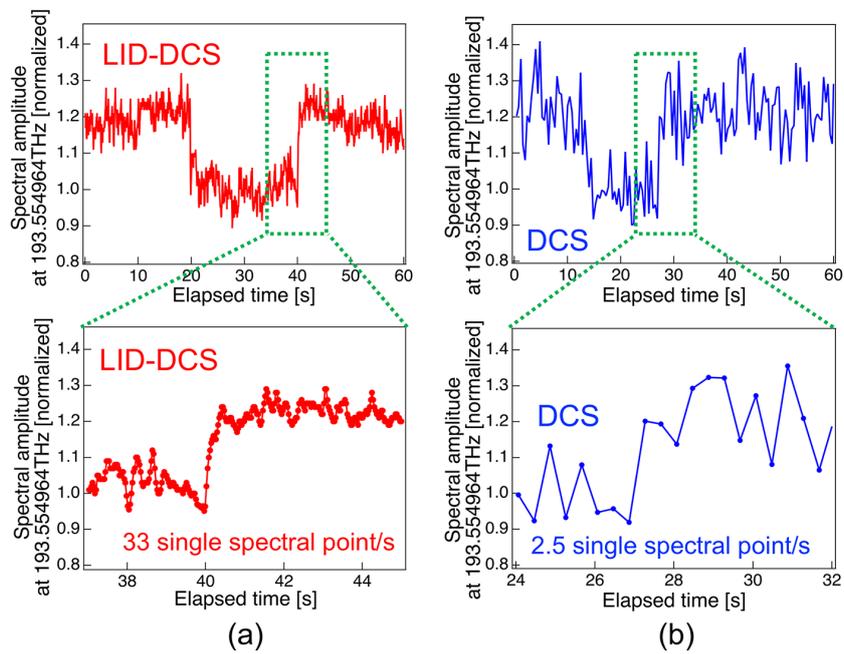

Fig. 3. Temporal response of spectral amplitude for (a) LID-DCS and (b) DCS when the optical beam of the signal light was chopped by a glass plate. Lower insets show the magnified temporal response of upper figures.

*3.3 Spectroscopy of cyanide gas ($H^{13}C^{14}N$)*

Finally, we demonstrated spectroscopy of cyanide gas ($H^{13}C^{14}N$) by LID-DCS and DCS. $H^{13}C^{14}N$ gas, contained in a gas cell (cell length = 15 cm , gas pressure = 25 Torr), was placed into an optical path of the signal light. We selected P(9) absorption line of $H^{13}C^{14}N$ gas (center frequency = 193.544907 THz, expected pressure-broadening linewidth = 2.25 GHz) for measurement. Red lines in Fig. 4(a) show a mode-resolved amplitude spectrum of OFC within the frequency range from 193.535 THz to 195.555 THz, measured by LID-DCS (optical spectral resolution = 3.24 MHz, LID time constant = 10 ms, number of signal accumulation = 4700, net acquisition time = 47 s/a single spectral point). We here extracted the RF comb mode at a frequency interval of 1 GHz across the P(9) absorption line by scanning the LID reference-frequency at a frequency interval of 2.12 kHz. This leads to the discrete mode distribution. Since the switching speed of LID frequency is sufficiently fast compared with the LID time constant, the acquisition time for the spectrum is limited by the LID time constant. Although the net acquisition time was set to be 47 s in this demonstration for high SNR spectrum, there is for further reduction of the acquisition time by reducing the number of signal accumulation. For comparison, we measured the same absorption line by DCS (optical spectral resolution = 10.0 MHz, number of consecutive interferograms = 10, time window size = 100 ns, sampling time interval = 21.2 fs, number of temporal data = 4,716,980, number of signal accumulation = 1000, acquisition time of interferogram = 47 ms/data, FFT calculation time = 3.5 s/data, net acquisition time = 50.5 s/a single spectral point), as shown in concentrated blue lines of Fig. 4(a). Envelopes of amplitude spectrum measured by LID-DCS and DCS were almost overlapped to each other as shown in the difference plot of them. To determine the center frequency and the linewidth of the measured P(9) absorption line measured by LID-DCS, we performed the curve-fitting analysis using the Gaussian function. Red plots and red line in Fig. 4(b) show the experimental data of the amplitude spectrum and the corresponding fitting result, respectively. For comparison, literature value of this absorption line position [30] is indicated as a green line in Fig. 4(b). The center frequency and the linewidth were determined to be 193.545100 THz and 2.33 GHz. The deviation of them from the expected values might be due to the instability or calibration error of the system used in the DCS, such as frequency counters, sinusoidal function generators, feedback controllers, and so on. However, these results indicate the correct frequency scale and high applicability for gas spectroscopy of LID-DCS.

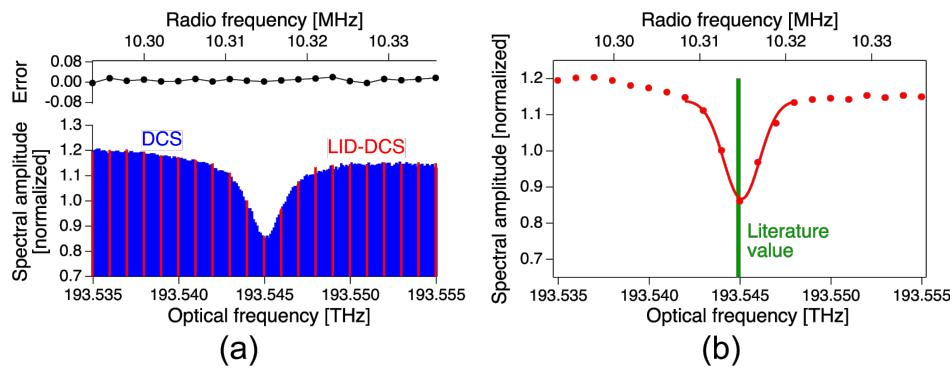

Fig. 4. (a) Amplitude spectrum around a P(9) absorption line of cyanide gas ($H^{13}C^{14}N$) measured by LID-DCS (red lines, sampling interval = 1 GHz) and DCS (blue lines,

sampling interval = 100 MHz), and its difference plot (upper). (b) Experimental data for the amplitude spectrum (red lines, sampling interval = 1 GHz) and its corresponding curve fitting analysis around the same absorption line. Green line shows a literature value of this absorption line position [30].

## 4. Discussion

One may have doubts about the advantage of LID-DCS compared with usual CW spectroscopy [30] because only a single OFC mode was extracted in this article. However, LID-DCS can rapidly select an arbitrary OFC mode within the broadband OFC spectrum, and frequency uncertainty of the selected OFC mode is always secured by a frequency standard. To realize the high spectral accuracy in CW spectroscopy as same accuracy as the LID-DCS, a CW laser must be stabilized to an OFC, indicating the fast wavelength tuning of the CW laser source would be difficult due to the requirement of a complicated method for the determination of absolute frequency, e.g., the tuning of the comb parameters such as $f_{rep}$, continuous wavelength sweeping of the CW laser source, and so on. Simultaneous acquisition of optical amplitude and phase will be another advantage although either the amplitude or phase signal was acquired in this article or the previous article [29]. Furthermore, a low frequency electrical noise can be suppressed in the LID-DCS owing to an inherent heterodyne detection mechanism of DCS, while the CW spectroscopy requires an additional intensity or frequency modulator to suppress the low frequency electrical noise. Multi-channel detection is an interesting option for the further extension of LID-DCS although a single-channel LID-DCS was used for gas spectroscopy in this article. The state-of-art multi-channel lock-in detection [29, 32, 33] will bring interesting options to LID-DCS. For example, multi-channel LID-DCS enables simultaneous monitoring of different gas samples without the time delay of FFT calculation in DCS. Also, simultaneous monitoring of different absorption lines in the same gas sample makes it possible to determine the gas temperature together with gas concentration [34]. These will be a powerful tool for analysis of combustion process in industry.

One may consider a possibility to further reduce the FFT calculation time in DCS by devising the data processing. FFT calculation time depends on the FFT calculation algorithm and PC speed. Although full spectrum of amplitude and phase in OFC was obtained by usual FFT calculation algorithm in this article, the specific FFT calculation algorithm suitable to obtain narrow-band spectral data acquisition may enables the further reduction of FFT calculation time in DCS. In this case, one have to compare the speed of electronics in LID-DCS and the speed of PC in DCS. This comparison will be our future work.

A combination of LID-DCS with imaging is another interesting extension. Single-pixel imaging (SPI) has attracted attention as a scan-less imaging scheme [35, 36] because SPI is particularly useful when the appropriate image detector is not readily available; for example, the RF interferogram measured in DCS is too fast for camera acquisition. In SPI, while the sample object is coded sequentially by a series of mask patterns generated by a spatial light modulator or a digital mirror device, the corresponding total light intensity passing through the sample object is measured as time-series data by a single-channel photodetector; then, the original two-dimensional image is reconstructed mathematically by the correlation calculation between the known coded masks and the measured time-series data of total light intensity. Recently, SPI was effectively combined with DCS for ultrahigh-resolution and ultrahigh-dense hyper-spectral imaging [19]. While this combination enables us to perform the scan-less imaging without the need for the image detector, FFT calculation time of interferogram on every coding pattern hampers the scan-less advantage of SPI and is a bottleneck for rapid imaging. If LID-DCS is combined with SPI in place of DCS, the image acquisition time will be largely reduced.

## 5. Conclusion

We demonstrated use of LID in DCS. This combination, LID-DCS, has potential to largely reduce the time spent for FFT calculation of a huge amount of temporal data because it depends on the frequency-domain measurement without the need for FFT calculation. Although the large amount of spectral data points available in the usual DCS are sometimes useful because of such as the application of spectral fitting to improve the accuracy of gas spectroscopy, LID-DCS benefits from the faster temporal response than usual DCS while maintaining the high resolution and accuracy comparable to usual DCS. Such characteristics of LID-DCS will be a powerful tool for monitoring of transient signal change, such as gas concentration measurement under air turbulence. Furthermore, options for multi-channel detection or imaging will expand the application fields of LID-DCS.

**Funding**

Exploratory Research for Advanced Technology (ERATO) MINOSHIMA Intelligent Optical Synthesizer Project (JPMJER1304), Japan Science and Technology Agency (JST), Japan, and Institute of Post-LED Photonics, Tokushima University, Japan.

**Acknowledgments**

The authors wish to acknowledge Ms. Shoko Lewis of Tokushima Univ., Japan, for her help in preparation of the manuscript.

# References

1. T. Udem, J. Reichert, R. Holzwarth, and T. W. Hänsch, "Accurate measurement of large optical frequency differences with a mode-locked laser," Opt. Lett. **24**, 881-883(1999).

2. M. Niering, R. Holzwarth, J. Reichert, P. Pokasov, T. Udem, M. Weitz, T. W. Hänsch, P. Lemonde, G. Santarelli, M. Abgrall, P. Laurent, C. Salomon, and A. Clairon, "Measurement of the hydrogen 1S-2S transition frequency by phase coherent comparison with a microwave cesium fountain clock," Phys. Rev. Lett. **84**, 5496-5499(2000).

3. T. Udem, R. Holzwarth, and T. W. Hänsch, "Optical frequency metrology," Nature **416**, 233–237 (2002).

4. P. R. Griffiths, Chemical infrared Fourier transform spectroscopy (Wiley, 1975).

5. P. Maslowski, K. F. Lee, A. C. Johansson, A. Khodabakhsh, G. Kowzan, L. Rutkowski, A. A. Mills, C. Mohr, J. Jiang, M. E. Fermann, and A. Foltynowicz, "Surpassing the path-limited resolution of Fourier-transform spectrometry with frequency combs," Phys. Rev. A **93**, 021802 (2016).

6. S. A. Diddams, L. Holloberg, and V. Mbele, "Molecular fingerprinting with the resolved modes of a femtosecond laser frequency comb," Nature **445**, 627-630(2007).

7. M. J. Thorpe, D. Balslev-Clausen, M. S. Kirchner, and J. Ye, "Cavity-enhanced optical frequency comb spectroscopy: application to human breath analysis," Opt. Express **16**, 2387-2397 (2008).

8. A. J. Fleisher, B. J. Bjork, T. Q. Bui, K. C. Cossel, M. Okumura, and J. Ye, "Mid-Infrared Time-Resolved Frequency Comb Spectroscopy of Transient Free Radicals", J. Phys. Chem. Lett. **5**, 2241-2246(2014).

9. M. Shirasaki, "Large angular dispersion by a virtually imaged phased array and its application to a wavelength demultiplexer," Opt. Lett. **21**, 366–368 (1996).

10. S. Xiao and A. M. Weiner, "2-D wavelength demultiplexer with potential for ≥1000 channels in the C-band," Opt. Express 12, 2895–2902 (2004).

11. S. Schiller, "Spectrometry with frequency combs," Opt. Lett. **27**, 766–768 (2002).

12. F. Keilmann, C. Gohle, and R. Holzwarth, "Time-domain mid-infrared frequency-comb spectrometer," Opt. Lett. **29**, 1542–1544 (2004).

13. T. Yasui, Y. Kabetani, E. Saneyoshi, S. Yokoyama, and T. Araki, "Terahertz frequency comb by multifrequency-heterodyning photoconductive detection for high-accuracy, high-resolution terahertz spectroscopy," Appl. Phys. Lett. **88**, 241104 (2006).

14. I. Coddington, N. Newbury, and W. Swann, "Dual-comb spectroscopy," Optica **3**, 414–426 (2016).

15. AM. Zolot, F. Giorgetta, E. Baumann, W. Swann, I. Coddington, and N. Newbury, "Broad-band frequency references in the near-infrared: accurate dual comb spectroscopy of methane and acetylene," J. Quant. Spectrosc. Radiat. Transfer **118**, 26–39 (2013).

16. Y. Shimizu, S. Okubo, A. Onae, K. M. T. Yamada, and H. Inaba, "Molecular gas thermometry on acetylene using dual-comb spectroscopy: analysis of rotational energy distribution," Appl. Phys. B **124**, 71 (2018).

17. Asahara, A. Nishiyama, S. Yoshida, K. Kondo, Y. Nakajima, and K. Minoshima, "Dual-comb spectroscopy for rapid characterization of complex optical properties of solids," Opt. Lett. **41**, 4971-4974(2016).

18. T. Minamikawa, Y. Hsieh, K. Shibuya, E. Hase, Y. Kaneoka, S. Okubo, H. Inaba, Y. Mizutani, H. Yamamoto, T. Iwata, and T. Yasui, "Dual-comb spectroscopic ellipsometry," Nat. Commun. **8**, 610 (2017).

19. K. Shibuya, T. Minamikawa, Y. Mizutani, H. Yamamoto, K. Minoshima, T. Yasui, and T. Iwata, "Scan-less hyperspectral dual-comb single-pixel-imaging in both amplitude and phase," Opt. Express **25**, 21947-21957(2017).

20. T. Ideguchi, S. Holzner, B. Bernhardt, G. Guelachvili, N. Picqué, and T. W. Hänsch, "Coherent Raman spectro-imaging with laser frequency combs," Nature **502**, 355–358 (2013).

21. S. Okubo, K. Iwakuni, H. Inaba, K. Hosaka, A. Onae, H. Sasada and F. L. Hong, "Ultra-broadband dual-comb spectroscopy across 1.0–1.9μm," Appl. Phys Express, **8**, 082402 (2015).

22. G. B. Rieker, F. R. Giorgetta, W. C. Swann, J. Kofler, A. M. Zolot, L. C. Sinclair, E. Baumann, C. Cromer, G. Petron, C. Sweeney, P. P. Tans, I. Coddington, and N. R. Newbury, "Frequency-comb-based remote sensing of green gases over kilometer air paths," Optica **1**, 290-298 (2014).

23. K. C. Cossel, E. M. Waxman, F. R. Giorgetta, M. Cermak, I. R. Coddington, D. Hesselius, S. Ruben, W. C. Swann, G.-W. Truong, G. B. Rieker, and N. R. Newbury, "Open-path dual-comb spectroscopy to an airborne retroreflector," Optica **4**, 724-728 (2014).

24. P. J. Schroeder, R. J. Wright, S. Coburn, B. Sodergren, K. C. Cossel, S. Droste, G. W. Truong, E. Baumann, F. R. Giorgetta, I. Coddington, N. R. Newbury, and G. B. Rieker, "Dual frequency comb laser absorption spectroscopy in a 16 MW gas turbine exhaust," Proceedings of the Combust. Inst. **36**, 4565-4573 (2017).

25. B. Berinhardt, E. Sorokin, P. Jacquet, R. Thon, T. Becker, I. T. Sorokina, N. Picque, and T. W. Hänsch, "Mid-infrared dual-comb spectroscopy with 2.4µm $Cr^{2+}$:ZnSe femtosecond lasers," Appl. Phys. B, **100**, 3-8 (2010).

26. T. Yasui, Y. Kabetani, E. Saneyoshi, S. Yokoyama, and T. Araki, "Terahertz frequency comb by multi-frequency-heterodyning photoconductive detection for high-accuracy, high-resolution terahertz spectroscopy," Appl. Phys. Lett. **88**, 241104 (2006).

27. Y.D. Hsieh, Y. Iyonaga, Y. Sakaguchi, S. Yokoyama, H. Inaba, K. Minoshima, F. Hindle, T. Araki, and T. Yasui, "Spectrally interleaved, comb-mode-resolved spectroscopy using swept dual terahertz combs," Sci. Rep. **4**, 3816 (2014).

28. S. Yokoyama, T. Yokoyama, Y. Hagihara, T. Araki, and T. Yasui, "A distance meter using a terahertz intermode beat in an optical frequency comb," Opt. Express **17**, 17324-17333 (2009).

29. R. Yang, F. Pollinge, K. Meiners-Hagen, M. Krystek, J. Tan, and H. Bosse, "Absolute distance measurement by dual-comb interferometry with multi-channel digital lock-in phase detection," Meas. Sci. Technol. **26**, 084001 (2015).

30. W. C. Swann and S. L. Gilbert, "Line centers, pressure shift, and pressure broadening of 1530-1560nm hydrogen cyanide wavelength calibration lines," J. Opt. Soc. Am. B, **22**, 1749-1756 (2005).

31. P. Werle, R. Mucke, and F. Slemr, "The limits of signal averaging in atmospheric trace-gas monitoring by tunable diode-laser absorption spectroscopy (TDLAS)," Appl. Phy. B **57**, 131-139 (1993).

32. N. Ishii, E. Tokunaga, S. Adachi, T. Kimura, H. Matsuda, and T. Kobayashi, "Optical frequency- and vibrational time-resolved two-dimensional spectroscopy by real-time impulsive resonant coherent Raman scattering in polydiacetylene," Phys. Rev. A **70**, 023811(2004).

33. P. Mao, Z. Wang, W. Dang, and Y. Wenga, "Multi-channel lock-in amplifier assisted femtosecond time-resolved fluorescence non-collinear optical parametric amplification spectroscopy with efficient rejection of superfluorescence background," Rev. Sci. Instrum. **86**, 123113 (2015).

34. Y. Deguchi, M. Noda, Y. Fukuda, Y. Ichinose, Y. Endo, M. Inada, Y. Abe, and S. Iwasaki, "Industrial applications of temperature and species concentration monitoring using laser diagnostics," Meas. Sci. Technol. **13**, R103-115 (2002).

35. W. K. Pratt, J. Kane, and H. C. Andrews, "Hadamard transform image coding," Proc. IEEE **57**, 58–68 (1969).

36. J. H. Shapiro, "Computational ghost imaging," Phys. Rev. A **78**, 061802 (2008).